# Integration of 5G with TSN as Prerequisite for a Highly Flexible Future Industrial Automation: Time Synchronization based on IEEE 802.1AS

Michael Gundall*, Christopher Huber*, Peter Rost†, Rüdiger Halfmann†, and Hans D. Schotten*‡
*German Research Center for Artificial Intelligence GmbH (DFKI), Kaiserslautern, Germany
†Nokia Bell Labs, Munich, Germany
‡Department of Electrical and Computer Engineering, Technische Universität Kaiserslautern, Kaiserslautern, Germany
Email: {michael.gundall, christopher.huber, hans_dieter.schotten}@dfki.de
{peter.m.rost, ruediger.halfmann}@nokia-bell-labs.com,

***Abstract*—Industry 4.0 brings up new types of use cases, whereby mobile use cases play a significant role. These use cases have stringent requirements on both automation and communication systems that cannot be achieved with recent shop floor technologies. Therefore, novel technologies such as IEEE time-sensitive networking (TSN) and Open Platform Communications Unified Architecture (OPC UA) are being introduced. In addition, for the realization of mobile use cases, wireline technologies cannot be used and have to be replaced by wireless connections, which have to meet the high demands of the industrial landscape. Here, 5th generation wireless communication system (5G) is seen as a promising candidate.**

**Especially encouraging and similarly challenging is the cooperative work of mobile robots, where particularly high demands on time synchronization arise. Therefore, this paper introduces a concept for the integration of TSN time synchronization (IEEE 802.1AS) conform with 5G to fulfill the requirements of these use cases. Furthermore, the paper describes a testbed for discrete manufacturing, consisting pre-dominantly of industrial equipment, in order to evaluate the presented approach.**

## I. Introduction

A high flexibility in the manufacturing process is envisioned to be most relevant for the factories of the future. The German Industry 4.0 vision describes better time, resource, and energy efficiency, optimization of product quality, and lot size one as the corresponding goals [1]. To achieve these goals, novel technologies must be introduced. In this context, technologies known from the information technology (IT), such as OS-level virtualization and Ethernet-based communication, are investigated and adapted for industrial purposes [2], [3].

Very characteristic for industrial applications are the stringent demands on cycle time, determinism, and availability that strongly differ from office applications. Consequently, the TSN working group is developing standards that define mechanisms for the time-sensitive transmission of data over deterministic Ethernet networks [4], [5].

In addition, the number of mobile applications such as mobile robotics (platooning, cooperative transport of goods, etc.) is steadily increasing. These use cases require high performance wireless communication systems. Here, the German Federal Ministry of Education and Research (BMBF) initiated the collaborative project Tactile Internet 4.0 (TACNET 4.0) [6], which describes use cases, concepts, and challenges to enable the development of efficient solutions that apply for both, discrete and process automation [7]. One of the major challenges is the convergence of wireline and wireless communication systems, as only few wireless connections have been deployed until now. Mobile radio communications, such as 5G, are well suited for this purpose but require novel concepts for a seamless integration into the operational technology (OT). Therefore, 3rd Generation Partnership Project (3GPP) Release 16 aims to support applications requiring deterministic or isochronous communication with high reliability and availability, such as TSN over 5G mobile networks. Furthermore, [8] proposed a concept that presents the 5G system to the TSN system like any other TSN-aware bridge.

In this context, we show by the example of time synchronization that the integration of 5G and TSN is advantageous. Therefore, the following contributions can be found in this paper:

- **Introduction of a concept for integrating TSN conform time synchronization (IEEE 802.1AS) with 5G in order to fulfill the requirements imposed by industrial mobile use cases.**
- **Performance evaluation of the proposed concept based on a discrete manufacturing testbed.**

Accordingly, the paper is structured as follows: Sec. II motivates our work on this topic, while Sec. III gives an overview about the related technologies. Details on our concept for the integration of IEEE 802.1AS and 5G are given in Sec. IV. This is followed by a performance evaluation of our approach based on a testbed for discrete manufacturing (Sec. V). Finally, the paper is concluded in Sec. VI.

## II. Target Use Cases

Many novel use cases are emerging in the context of Industry 4.0 [7]. These use cases are essential in order to ensure the required flexibility of a smart manufacturing . The realization is particularly demanding for mobile use cases as they require wireless communication links due to their movement. Very challenging are those use cases where several mobile devices have a collaborative task, as here the most precise time and state synchronization is required.

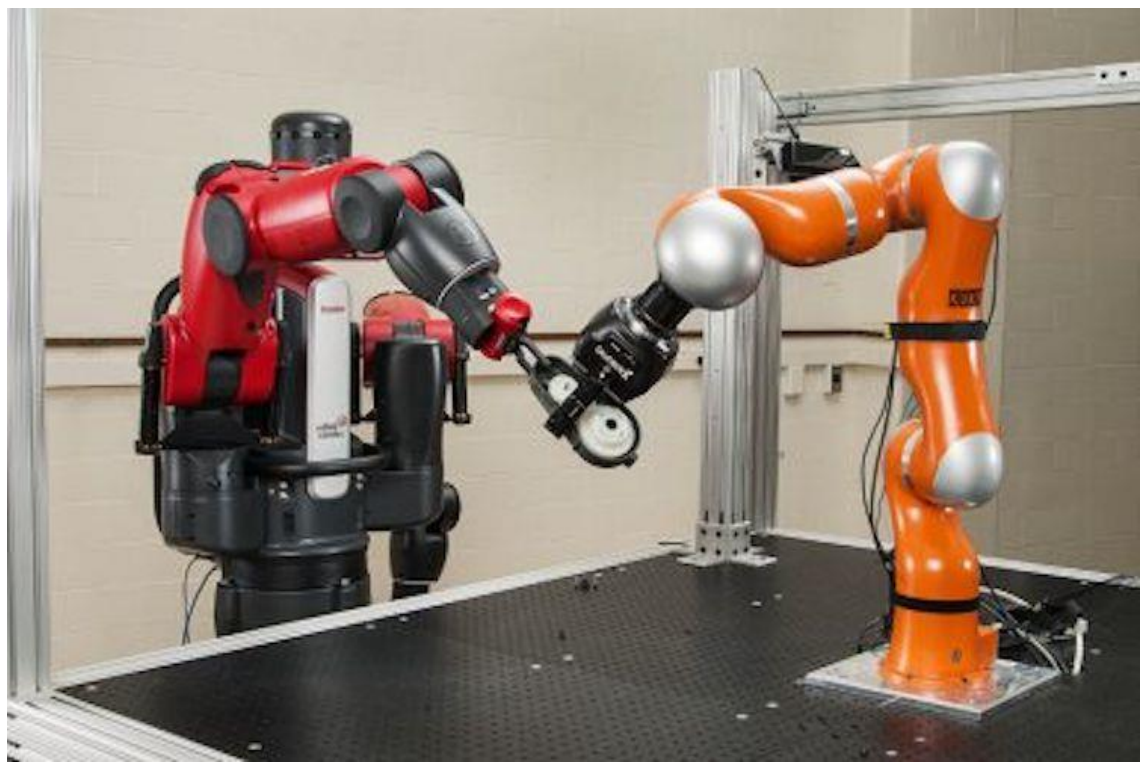

Figure 1. Collaborative robots [9]

Fig. 1 shows two robots working together on the same workpiece. Even if only one of the two robots, e.g. the red one, is supposed to be mobile, both a wireless connection and a precise time synchronization are required, whereby a better synchronicity leads to a faster interaction of the robots and thus to a higher productivity. Typically, this type of use cases require a synchronicity of $<$1ms [7].

## III. Background

To meet the stringent demands of upcoming use cases, several novel technologies are being introduced. This section provides an overview of the most relevant technologies that have been used for our work.

### A. IEEE Time-Sensitive Networking (TSN)

TSN is a set of standards being developed by the Time-Sensitive Networking task group of the IEEE 802.1 working group [4]. The recently developed standards extend the capabilities of Audio Video Bridging (AVB) and define mechanisms for the time-sensitive transmission of data over deterministic Ethernet networks with respect to guaranteed end-to-end (E2E) latencies, reliability, and fault-tolerance [5]. The use of IEEE 802 Ethernet in industrial applications that fulfill the requirements of industrial environments can also replace vendor-specific real-time solutions.

Furthermore, TSN consists of a number of different standardization documents, where IEEE 802.1Qav, IEEE 802.1AS, and IEEE 802.1Qat, describe different aspects of the technology and IEEE 802.1BA them all into context. Since our focus is on time synchronization for industrial applications, only the IEEE 802.1AS standard will be relevant for this paper.

### B. OPC Unified Architecture (OPC UA)

OPC UA is the platform independent successor of the OPC standard, developed by the OPC Foundation [10]. The general aim of OPC UA is the secure, easy and platform independent exchange of information between industrial appliances. Therefore, it provides several communication protocols for the data transport.

Since control commands, process alarms, and other events occur in an acyclic fashion, and the loss of one of those data packets would have severe consequences, the TCP-based server-client OPC UA communication protocol (UATCP) is well suited for supervisory control and data acquisition (SCADA) systems, but is not suited for closed-loop control, that require a cyclic and real-time data transmission. Therefore, part 14 of the OPC UA specifications adds the Publish Subscribe (PubSub) pattern allowing many subscribes to register for a specific content [11]. For the message distribution both broker-based protocols, in particular message queuing telemetry transport (MQTT) and advanced message queuing protocol (AMQP), and UADP, a custom UDP-based distribution based on the IP standard for multicasting has been defined. Due to the advantages to send real-time messages on the field level directly on the data link layer, part 14 defines the transport of PubSub messages based on Ethernet frames.

Since TSN aims at a time-sensitive transmission of data over deterministic Ethernet networks, the combination of TSN and OPC UA PubSub has already been discussed [12].

### C. 3GPP 5th Generation Wireless Communication System (5G)

5G is being developed by the 3GPP and signifies a major step forward in the capabilities of mobile networks. This wireless technology will advance traditional mobile broadband to a new level in terms of data rates, capacity, and availability. The most significant features can be divided into three parts: (1) massive Internet of Things (IoT), which is also known as massive machine-type communications (mMTC), (2) ultra-reliable low-latency communication (URLLC), and (3) enhanced mobile broadband (eMBB). However, new findings show that not all use cases can be assigned to these categories. Therefore, an additional category called “NR-Lite” is planned for Release 17, which will be available until the end of 2022 [13].

Alongside the functional enhancements, non-public networks (NPNs) emerged, making it possible for organizations or groups of organizations to operate their own (private) mobile network [14]. NPNs have several advantages over public networks in terms of performance, privacy, and security. By guaranteeing the requested quality of service (QoS), 5G is able to support services such as industrial IoT (IIoT) and mission critical communications, such as the cooperative work of mobile robots, where a precise time synchronization is required [15]. Therefore, 3GPP Release 16 describes mechanisms for the distribution of the TSN clock and time-stamping, according to IEEE 802.1AS [16]. On receiving a Downlink (DL) generalized Precision Time Protocol (gPTP) message the Network-Side TSN Translator (NW-TT) makes an ingress timestamping (TSi) for every gPTP event message and uses the cumulative rateRatio obtained within the payload of the gPTP message in order to calculate the link delay from the upstream TSN node (gPTP entity) expressed in TSN grandmaster (GM) time [17]. Afterwards the NW-TT calculates the new cumulative rateRatio as well as modifying the gPTP message payload. Besides adding the link delay from the upstream TSN node in TSN GM time to the correction field, the cumuluative rateRatio received from the upstream TSN node is replaced with the new cumulative rateRatio and a TSi in the suffix field of the gPTP packet is added. Then the gPTP message is being forwarded by the User Plane Function (UPF) via user plane. Therefore, all gPTP messages are transmitted on a QoS flow that meets the residence time upper bound requirement, which is specified in IEEE 802.1AS. After a user equipment (UE) receives the gPTP messages, it forwards them to the Device-Side TSN Translator (DS-TT), which creates an egress timestamping (TSe) for the gPTP event messages for external TSN working domains. The residence time spent wihin the 5G system (5GS) for this gPTP message is calculated by the difference between TSi and TSe. The DS-TT converts the residence time spent within the 5GS in TSN GM time with the rateRatio provided by the gPTP message. It also adds the calculated residence time and removes the TSi from the payload suffix field of the gPTP message sent to the downstream TSN node [18].

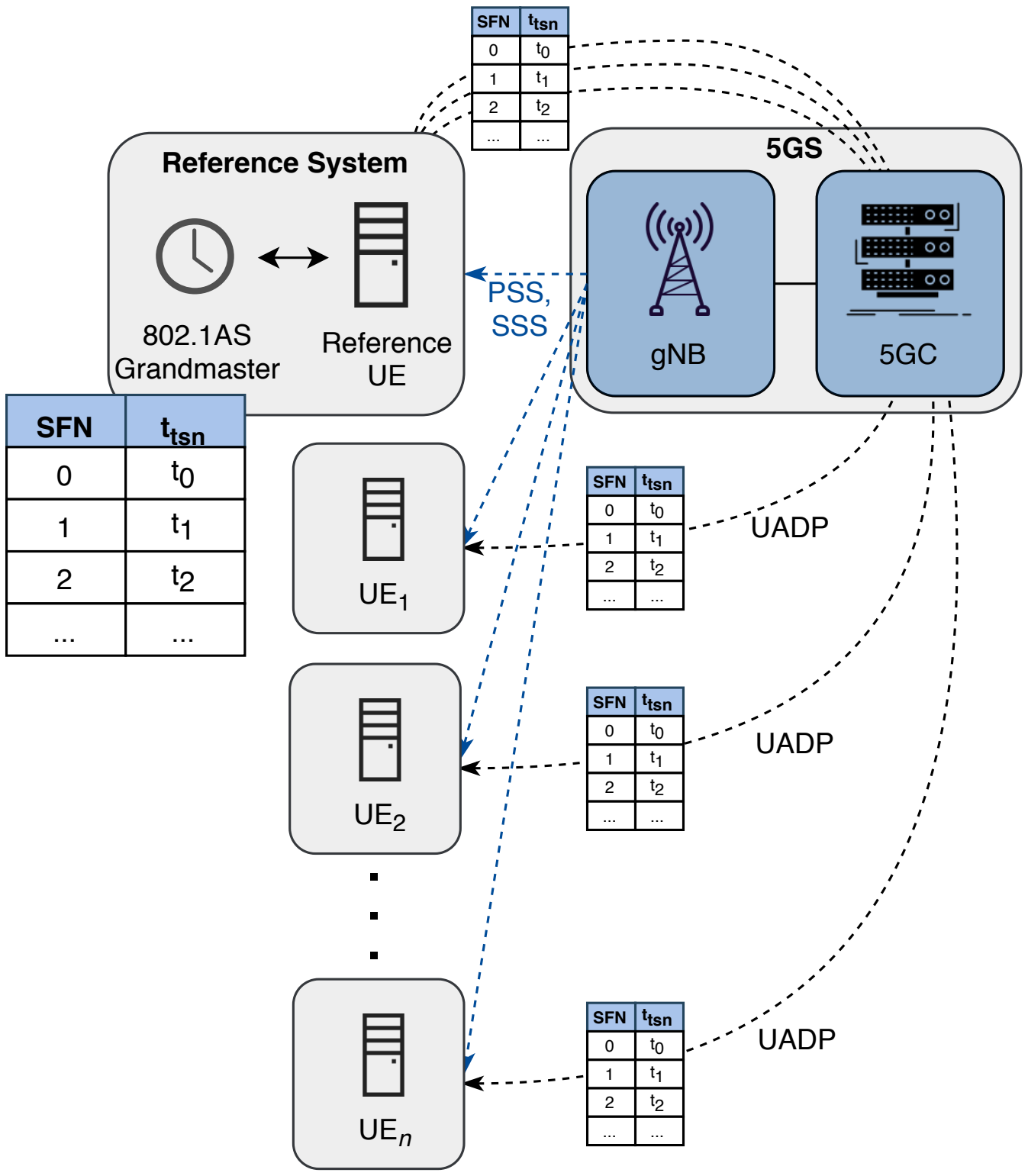

Figure 2. Concept for the distribution of the TSN time in the 3GPP network.

## IV. Concept for Integration of 5G with IEEE 802.1AS

The seamless integration of 5G with technologies of the OT leads to several challenges that need to be solved. In the previous section, the mechanism for time synchronization planned for 3GPP 5G Release 16 was already explained. However, as there is currently no 3GPP Release 16 hardware available to evaluate these mechanisms, we have developed an alternative solution that is also compatible with 4G, as no changes have been made to the base station or core network. Therefore, this section proposes a concept for the application of IEEE 802.1AS-based time synchronization within mobile radio networks.

The concept, which is shown in Fig. 2 consists of several components. A complete 5GS, which includes a 5G base station (gNB) and a 5G core network (5GC), and multiple UEs, whereby one of them is named “Reference UE” and is part of the Reference System. What makes this UE special is that it is connected to the wireline TSN network and can consequently not be mobile. In addition, this UE is synchronized with the TSN time and must support IEEE 802.1AS. This synchronization is done by the GM, which can be any TSN device.

Furthermore, the 5GS has its own synchronization mechanism that works as follows. In order to synchronize the radio access network (RAN), each gNB synchronizes its connected UEs, by using primary synchronization signal (PSS) and secondary synchronization signal (SSS). Furthermore, those signals are needed to help the UEs to detect the cell identity and to get radio frame boundary. The 5G synchronization is based on the beam management operations, which allows the physical link connection between a mobile UEs and a gNB. Furthermore it enables handover, radio link failure recovery procedures, beam tracking and beam adaptation for already connected devices. By utilizing PSS, SSS and specific synchronization algorithms, the UEs can approximate and adjust the frequency and time offsets. Once the PSS has been decoded, the UEs can identify the cell ID sector. Then by using the identified cell ID sector, the UEs can decode the SSS and identify the cell ID group [19]. These procedures are used to detect the sample timing within the full frame. Apart from this the UEs should identify the system frame number (SFN), which is a part of the master information block (MIB) and the Physical Broadcast Channel (PBCH) transport block.

5G has a 10 bit SFN starting from 0 to 1023 that increases every 10 ms. The MIB carries 6 most significant bit (MSB) and the PBCH transport block contains the last 4 least significant bit (LSB). In addition to the SFN a subframe number exists beginning from 0 to 9 that ticks every 1 ms. When the maximum value is reached, it resets and increases the SFN by 1. When the SFN reaches the maximum value, it starts over again. To preserve a stable connection the UE and gNB need to maintain the synchronization on SFN and subframe number at any time during the communication period.

For this reason, the given synchronization between the gNB and the connected UEs can be used and only the offset of the corresponding TSN time needs to be identified. Fig. 3 shows the flowchart for the Reference System. Thus, each incoming SFN is paired with the current time of the Reference UE and is sent to each UE that subscribes to this service. By using OPC UA PubSub for the distribution, it is possible to synchronize as many UEs as are connected to the gNB, with the message layers shown in Fig. 4. The transport protocol used is UDP in combination with multicast, which means that each of the UEs that have joined the multicast group receives the subscribed messages. If necessary, the transport protocol can also be changed from broker-less to broker-based, e.g. MQTT or AMQP. It is also possible to distribute the messages via multicast based on layer 2. In addition there is the so-called OPC UA *NetworkMessage* which forms the payload of the UDP datagram, each *NetworkMessage* having the OPC UA specific header and footer and containing one or more *DataSetMessages*, which in turn have so-called *DataSetMessage* fields. In our case, the *NetworkMessage* contains only one *DataSetMessage*, which consists of its header and the following two *DataSetMessage* fields: $SFN$, $t_{TSN}[SFN]$.

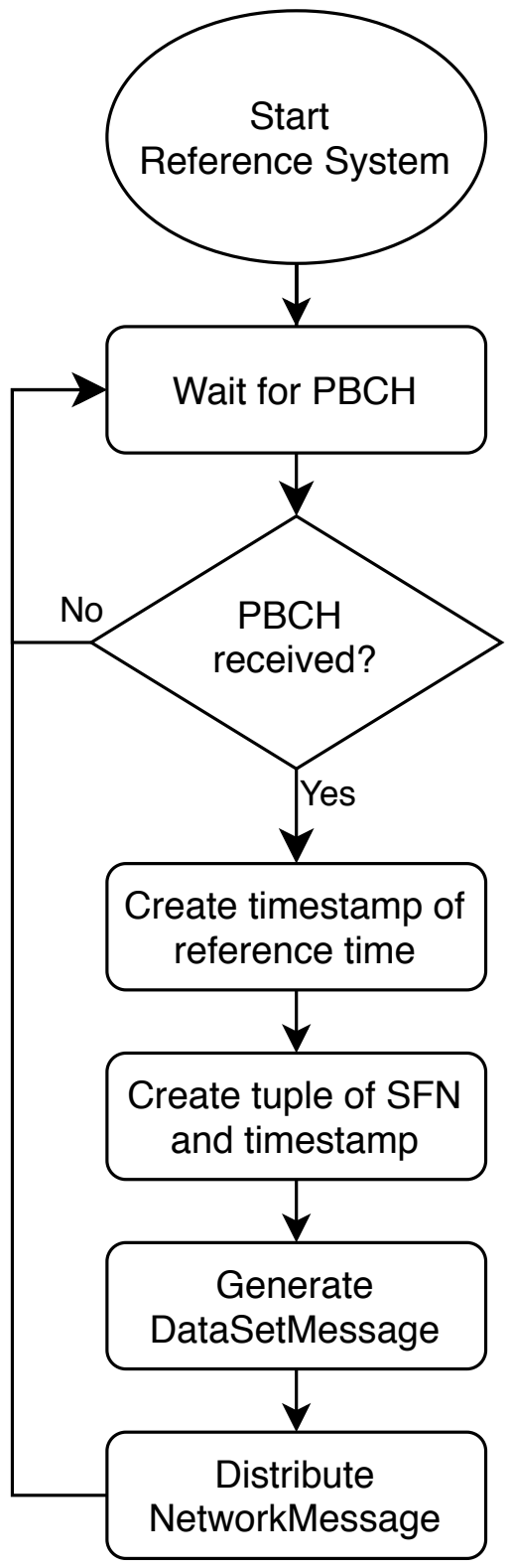

Figure 3. Flow diagram of the Reference System

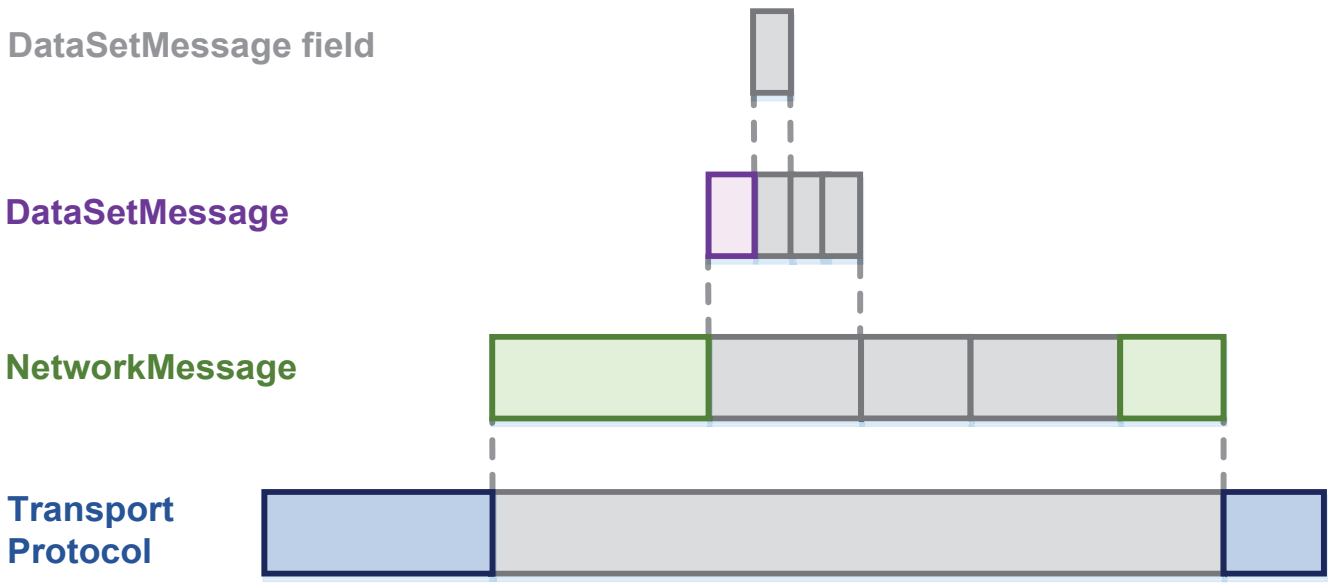

Figure 4. OPC UA PubSub message layers [11]

The second workflow, which is the same for all other UEs, is shown in Fig. 5. Here each UE derives the tuples of its specific time for each incoming SFN. These tuples are used alongside the received tuples to calculate the offset and adjust the local clock accordingly. The formula for adjusting the local clock of the mobile UEs is as follows, where $t_{TSN}[SFN]$ is the time of the Reference System for a specific SFN, $t_{UE}[SFN]$ is the local time of the UE that gets synchronized for the given SFN, and $t_{UE}[current]$ is the current time of the UE:

$$t_{TSN} = t_{TSN}[SFN] - t_{UE}[SFN] + t_{UE}[current] \quad (1)$$

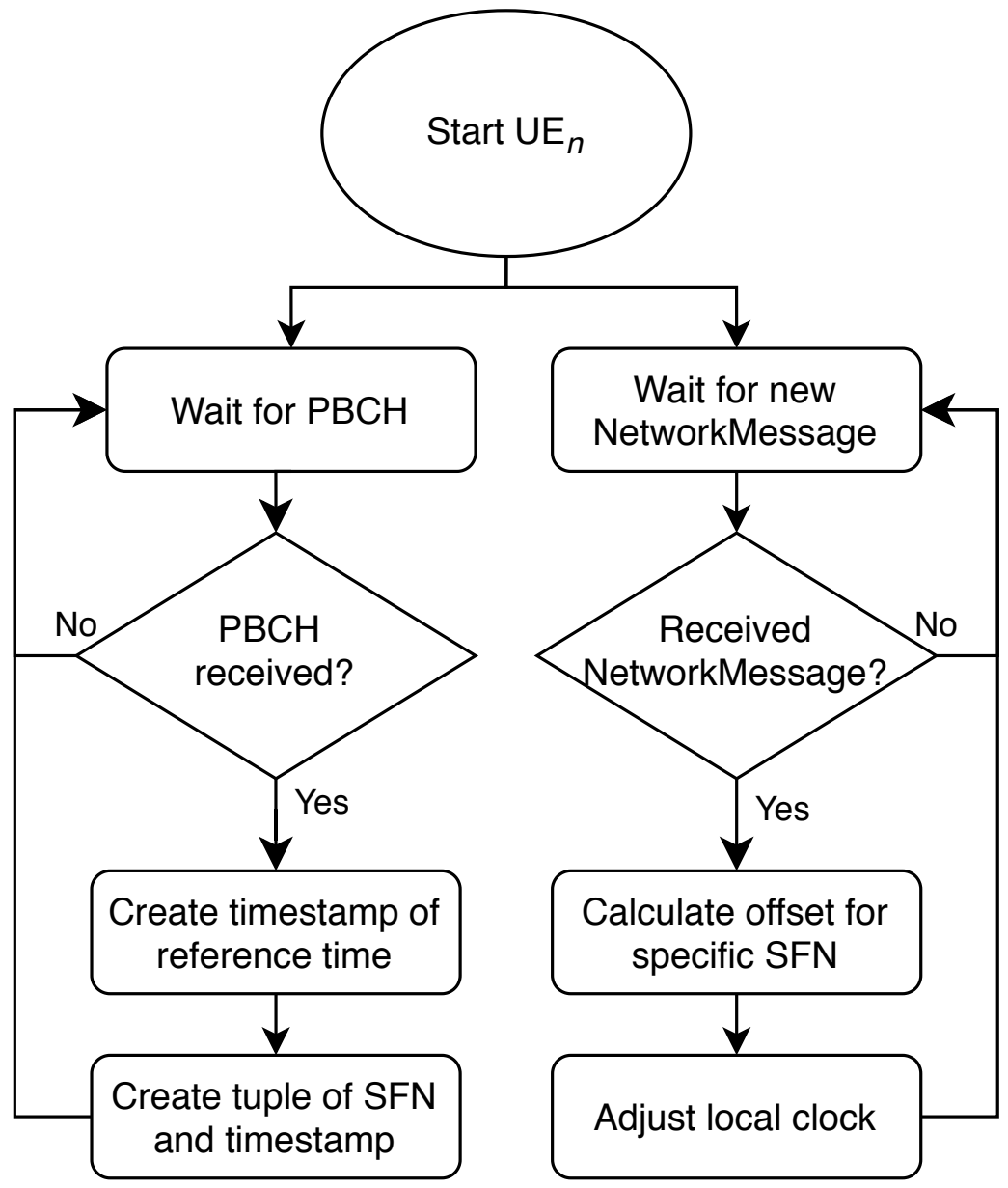

Figure 5. Flow diagram of a UE that gets synchronized by the Reference System

## V. Testbed & Validation

This section aims to demonstrate the validity of the proposed concept. Therefore, Sec. V-A describes the hardware and software components that are required to verify the concept presented in Sec. IV. Additionally, Sec. V-B introduces a discrete automation demonstrator based on the use case group that was discussed in Sec. II.

### A. Hardware & Software Setup for the Time Synchronization

Table I
Hardware configurations

| ***Equipment*** | ***QTY*** | ***Specification*** |
|---|---|---|
| Mini PC | 4 | Intel Core i7-8809G, 2x16 GB DDR4, Intel i210-AT & i219-LM Gigbabit NICs, Ubuntu 18.04.3 LTS 64-bit, Linux 4.18.0-18-lowlatency |
| Software Defined Radio | 3 | Ettus Research B210, RF 70MHz-6GHz, 2x2 MIMO, Spartan 6 FPGA |
| TSN Evaluation Kit | 1 | RAPID-TSNEK-V0001, IEEE 802.1AS |

In the following section the required hard- and software components for the setup of the concept are described in detail. Moreover, specifications for each of the used hardware components can be found in Tab. I.

*1) TSN Evaluation Kit:* The TSN Evaluation Kit is configured to provide TSN gateway functionality to ensure a quick evaluation of TSN features. It is possible to use the fido5000 real-time Ethernet multi-protocol (REM) switch chip for providing a TSN solution in the application of the product. By using the TSN gateway functionality, a non-TSN device can participate in a TSN network without implementing TSN-specific features natively. In addition, the TSN Evaluation Kit supports the IEEE 802.1AS and 802.1AS-REV specifications and can consequently serve as GM for other TSN devices.

*2) Linux PTP:* Linux PTP[1] is a free and open source software Precision Time Protocol (PTP) implementation that complies with the IEEE 1588 standard. This implementation is one of the most frequently used. Besides aiming to provide a robust implementation of the standard Linux PTP tries to make use of the most relevant and modern application programming interfaces (APIs) offered by the Linux kernel. The Linux PTP project provides several executables to run two-stage synchronization mechanism. The one which was used in our testbed is *ptp4l*.

The *ptp4l* tool synchronizes the PTP hardware clock with the master clock in the network. If there is no PTP hardware clock, it automatically synchronizes the system clock with a master clock using software timestamps. As extension, the tool supports the IEEE 802.1AS specification for TSN end stations, by using the gPTP configuration file, which modifies the default procedure of the executable.

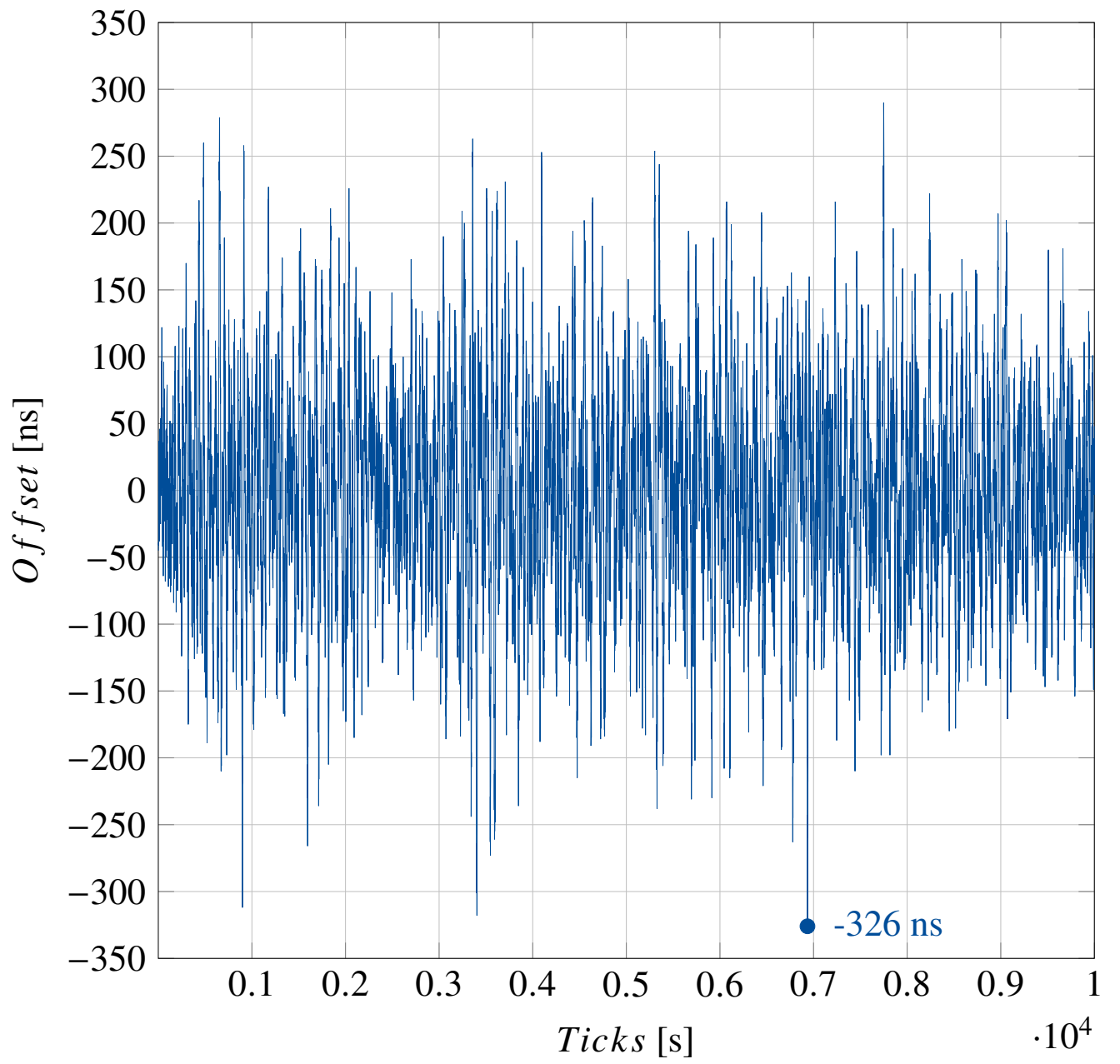

Figure 6. Synchronization accuracy between TSN Evaluation Kit and Intel NUC mini PC for a sync interval of 31.25ms ($2^{-5}$s).

As shown in Fig. 6 a synchronicity of ± 350ns between TSN Evaluation Kit and one of the mini personal computers (PCs) can be reached, by using the minimum sync interval of 31.25ms ($2^{-5}$s).

*3) OpenAirInterface (OAI):* OAI is an open experimentation and prototyping platform created by EURECOM to enable innovation in the area of mobile and wireless networking and communications. Furthermore, this platform is compliant with the 3GPP 4G and 5G standards. However, as there is not yet a stable 5G implementation, the 4G implementation is used. Here, OAI provides an implementation for 4G base station (eNB), UE, and evolved packet core (EPC) operating on general purpose computing platforms (x86) together with commercially available Software Defined Radio (SDR) such as Ettus Research Universal Software Radio Peripheral (USRP). Using this platform, it is possible to build and adapt 4G networks on PCs and connect them to commercial UEs such as smartphones, or to software-based UEs. The software provides a suitable development environment that allows the entire network to be monitored in real time. There are also other built-in tools such as highly realistic emulation modes, debugging tools, protocol analyzer and a logging system for all layers and channels. In our testbed the version of OAI we use for the eNB is the master branch with the commit id 82e5410 and for the UE master branch with the commit id 82e5410.

*4) USRP B210 (Software Defined Radios):* For realistic experimentation and validation, a software-defined radio module from Ettus Research, the USRP B210[2], is utilized in this testbed. It is a Multiple Input Multiple Output (MIMO) board with an integrated Spartan 6 Field Programmable Gate Array (FPGA), designed for continuous frequency coverage from 70 MHz to 6 GHz, due to the AD9361 analog transceiver. Additionally, it combines access to a channel with a large bandwidth of 56 MHz. Taking this specification into account, it is possible to use this device in the Industrial, Scientific and Medical (ISM) and evolved UMTS Terrestrial Radio Access (E-UTRA) frequency band.

### *B. Discrete Automation Demonstrator*

To evaluate the performance of the proposed concept, a discrete automation demonstrator is proposed, based on the use case presented in Sec. II. To make measurements as simple and error-free as possible and to visualize the validity of our concept, the basis of our demonstrator is the simultaneous one-dimensional motion of two carriages on linear axes, as it is shown in Fig.7.

The idea is that both carriages move together from a common start position p1 to a defined end point p2. If both the accelerations and the speeds of the two carriages are the same, the mechanical synchronicity can be determined by measuring the position difference, the greatest value being when the maximum speed is reached (see Eq. 2 & Eq. 3). If the speeds of the two carriages are known, the time delay $\Delta t$ with which the two carriages started their movement can be determined. This corresponds to the time synchronicity.

$$\Delta s(t) = s_1(t) - s_2(t) \tag{2}$$

[1]Further information: http://linuxptp.sourceforge.net/

[2]Further information: https://www.ettus.com/all-products/ub210-kit/

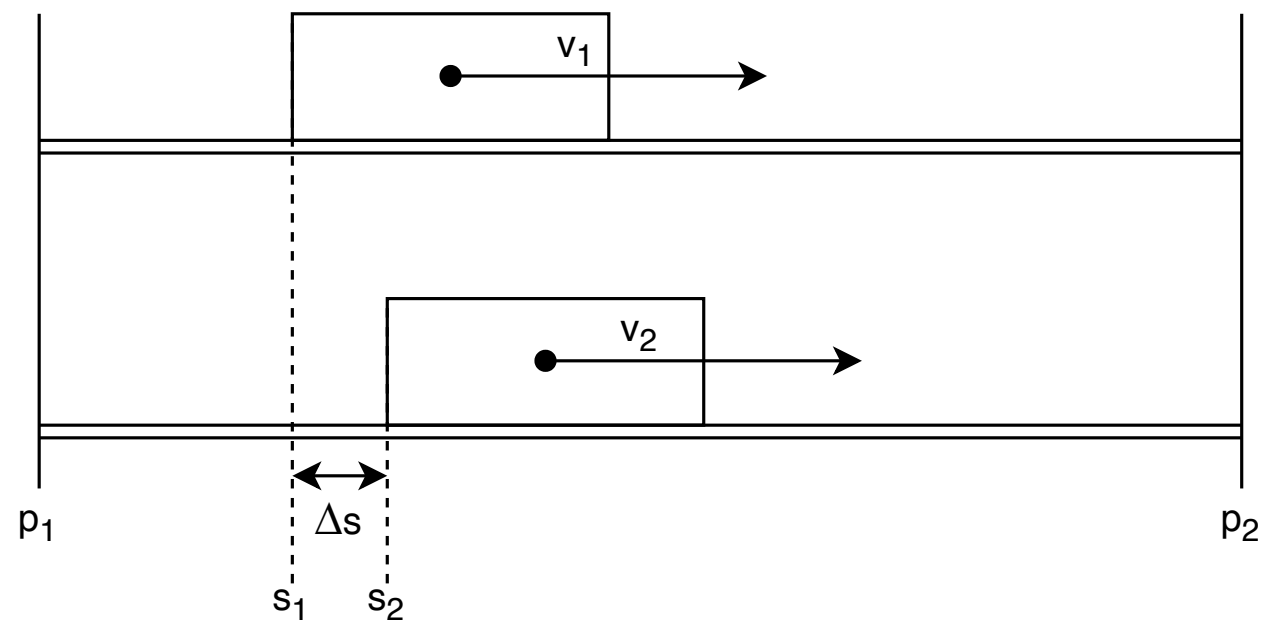

Figure 7. Concept for the identification of the mechanical offset and time synchronisation.

$$\Delta t_{max} = \frac{\Delta s_{max}}{v_{max}} \tag{3}$$

In order to validate our concept, it must be demonstrated that the required synchronicity of <1ms can be achieved. This requires both a high speed of the carriages and a high precision of the measurements. Due to the required dynamics two toothed belt axes of type ZLW-1040-S[3] in combination with two brushless DC (BLDC) motors of type MOT-EC NEMA24[4] as shown in Fig. 8, are used. By combining these two components, the carriages can reach a maximum speed of $|v_{max}| = 4\frac{m}{s}$ and an acceleration of $a_{max} = \pm 30\frac{m}{s^2}$. The latter ensures that the maximum speed is given over a major part of the 1m stroke. Furthermore, the BLDC motors require a 48V rotary field. Therefore, two motor controllers called dryve D1[5] are used, one connected to each motor and containing the necessary power electronics to invert the 48V input signal of the controller. In addition, the motor controllers control the target positions using the incremental encoders built into the motors. Several GIPOs at the input of the motor controllers can be used to specify both setpoints for the target positions and the start of the movement. Since we use mini PCs as UEs, which do not provide such interfaces, they are connected via USB to USB breakout modules, which are also shown in Fig. 8. Since the negative acceleration, which slows down the carriages, causes the motors to run in generator operation and thus increases the DC link voltage of the inverter, load resistors of 3.3Ω each are connected to the power inverters of the motor controllers.

Two high-precision sensors of type OM70-L1000.HV0500.VI[6] have been installed to measure the position of each slide. In their high-precision mode, in which a Kalman filter uses a time series of several values to reduce the noise, the resolution is about 3-63$\mu$m depending on the position of the carriages. This precision is sufficient to validate our concept. To analyze the measured values, a Programmable Logic Controller (PLC) of type S7-1512SP[7] along with two high-speed analog input units[8] that are connected to the current outputs (4-20mA) of the sensors are used and provide a resolution of 12 bit.

In addition to the measurement, an optical visualization is offered to demonstrate that the specified synchronicity is achieved. For this purpose, a laser pointer is located on one carriage and a screen on the other carriage on which the laser beam is displayed. Furthermore, an obstacle with a small hole is placed in front of the shield of the second carriage. When the required synchronicity is reached, the laser beam passes through the hole of the obstacle during the entire movement. Otherwise it is absorbed by the obstacle and no point on the shield is visible.

For the validation we perform three kinds of measurements: (1) synchronizing two mini PCs with the standard IEEE 802.1AS (gPTP) synchronization mechanism without any wireless communication, (2) tunneling the IEEE 1588 (PTP) messages on UDP basis over the 4G system, and (3) applying our concept for the integration of 3GPP 5G with IEEE 802.1AS. The results are shown in Fig. 9.

In the first setup, both mini PCs are wireline connected to the TSN Evaluation Kit and are synchronized by the GM. Fig. 9a shows that with this setup a mechanical synchronicity of ±4.7mm can be achieved.

Next, the transmission of PTP over 4G will be analyzed (see Fig. 9b). Although there are sections where there is similar synchronicity to Test (1), there are data points that are in the range of ±81.1mm. This is due to the fact that the line delay, which is the basis for the offset calculation of PTP and gPTP, varies with each transmission of wireless communication signals. However, a deviation of more than 8cm, which corresponds to a time synchronization accuracy of less than 2ms, between the two positions is not acceptable for industrial environments.

Finally, the results of Test (3) are evaluated, where our concept for the integration of IEEE 802.1AS with 3GPP has been applied. Here, Fig. 9c shows, that the maximum position offset $|\Delta s_{max}| = 2.2$mm. If Eq. 3 is applied and the maximum position offset is divided by the maximum speed, the maximum time delay of the start of the movement of both carriages, which corresponds to the mechanical synchronicity of the system, is 0.55 ms. This means that our concept in combination with the hardware of the demonstrator fulfills the desired synchronicity of <1ms. Even if the measured data indicate that our concept is even better than the TSN synchronicity of Test (1), this is due to the tolerances of the mechanical components of the hardware setup. It is clear that wired synchronization is more accurate than wireless synchronization. However, the results show that for our discrete automation demonstrator, which already has quite high dynamics, the same results can be achieved with a 3GPP 4G system compared to a wired system synchronized with IEEE 802.1AS.

As a summary of the measurement results Fig. 10 shows all curves in a single diagram. Here, it is visible the improvement by our concept. In fact, using our concept increases synchronicity by 360% when considering the maximum offset of the carriage positions.

[3]Further information: https://www.igus.eu/info/linear-guides-zlw-1040

[4]Further information: https://www.igus.eu/product/19073?L=en

[5]Further information: https://www.igus.eu/product/17827

[6]further information: https://www.baumer.com/de/en/product-overview/ Distance Measurement/Laser Distance Sensors/High Power/ large measuring distances up to-1500-mm/om70-l1000-hv0500-vi/p/38587

[7]Additional information: https://support.industry.siemens.com/cs/pd/578293

[8]Additional information: https://support.industry.siemens.com/cs/pd/129051

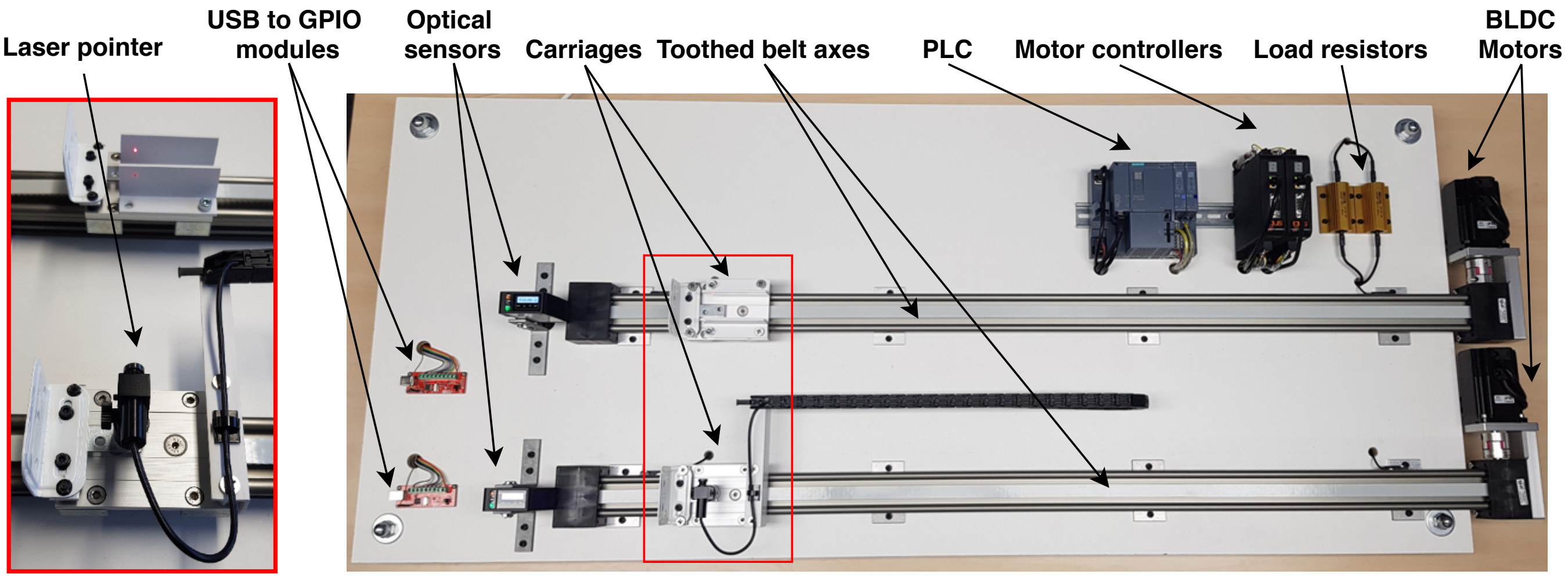

Figure 8. Hardware setup of the discrete automation demonstrator

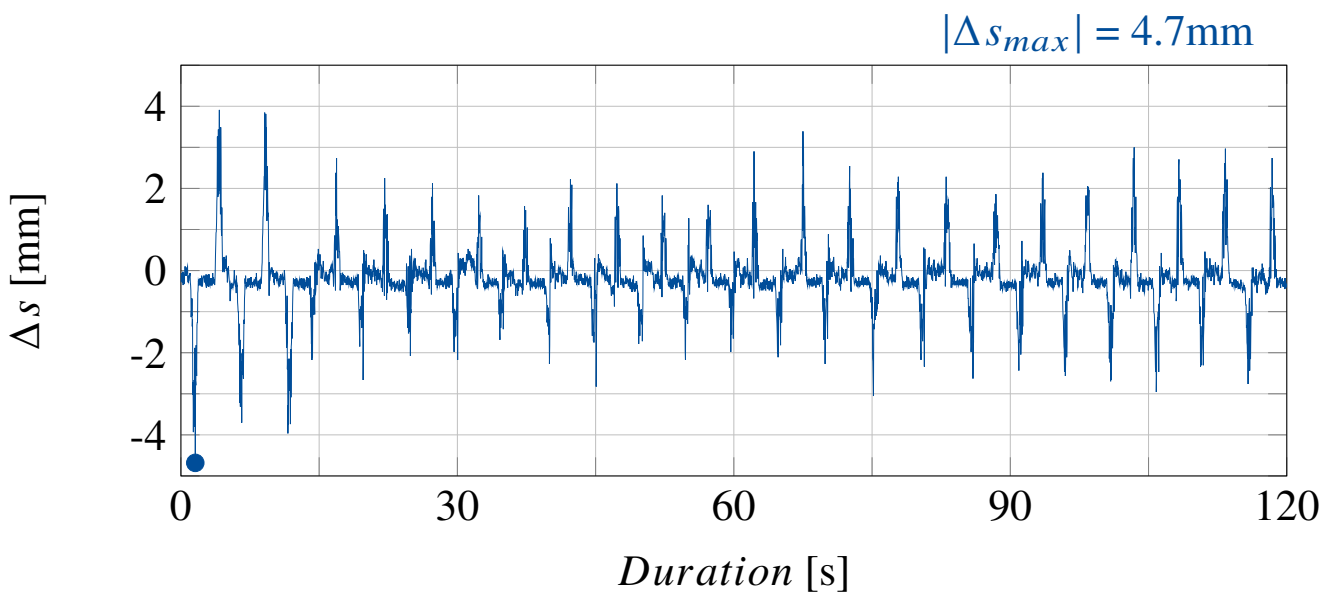

(a) IEEE 802.1AS (gPTP) based on wireline communication

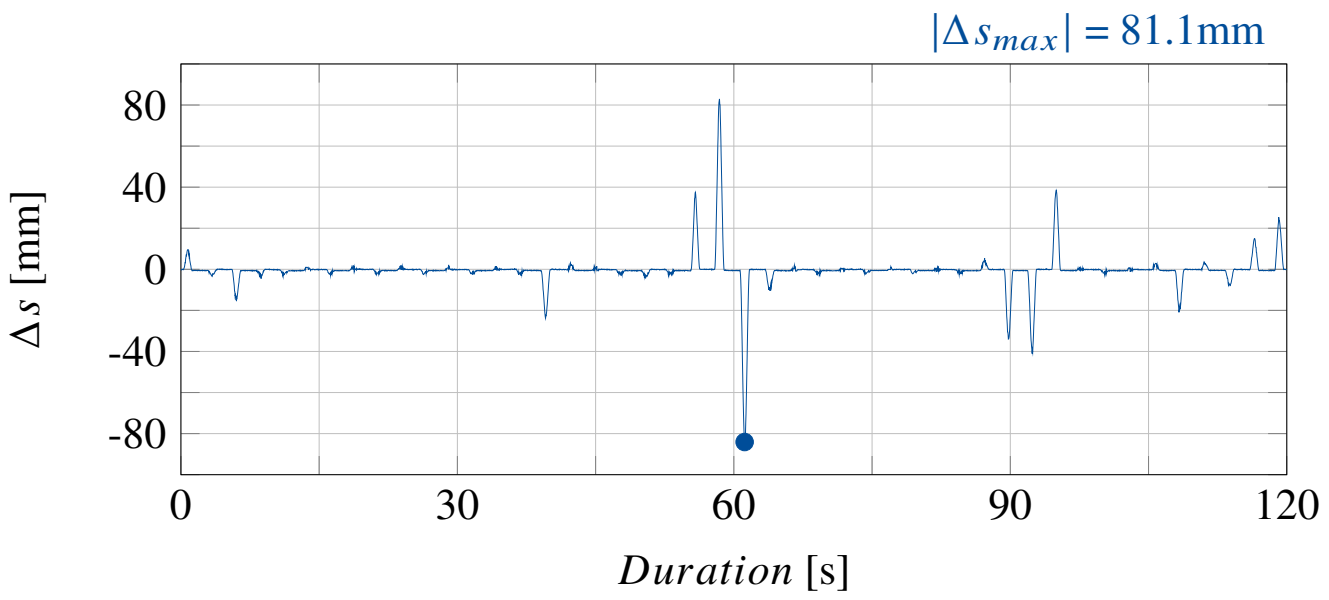

(b) Tunneling of IEEE 1588 (PTP) over a 3GPP 4G system

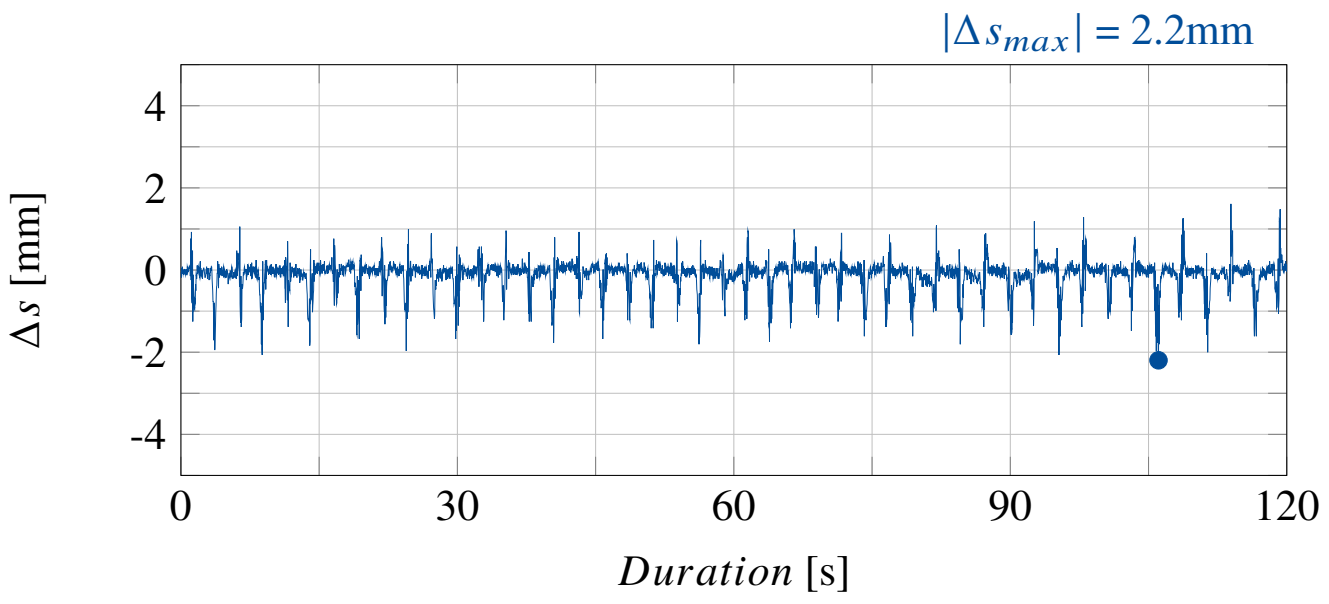

(c) Integration of 3GPP 5G with IEEE 802.1AS based on a 4G system

Figure 9. Readings of the measurements for the evaluation of our concept for the integration of 3GPP 5G with IEEE 802.1AS compared to a wireline setup and a state of the art solution.

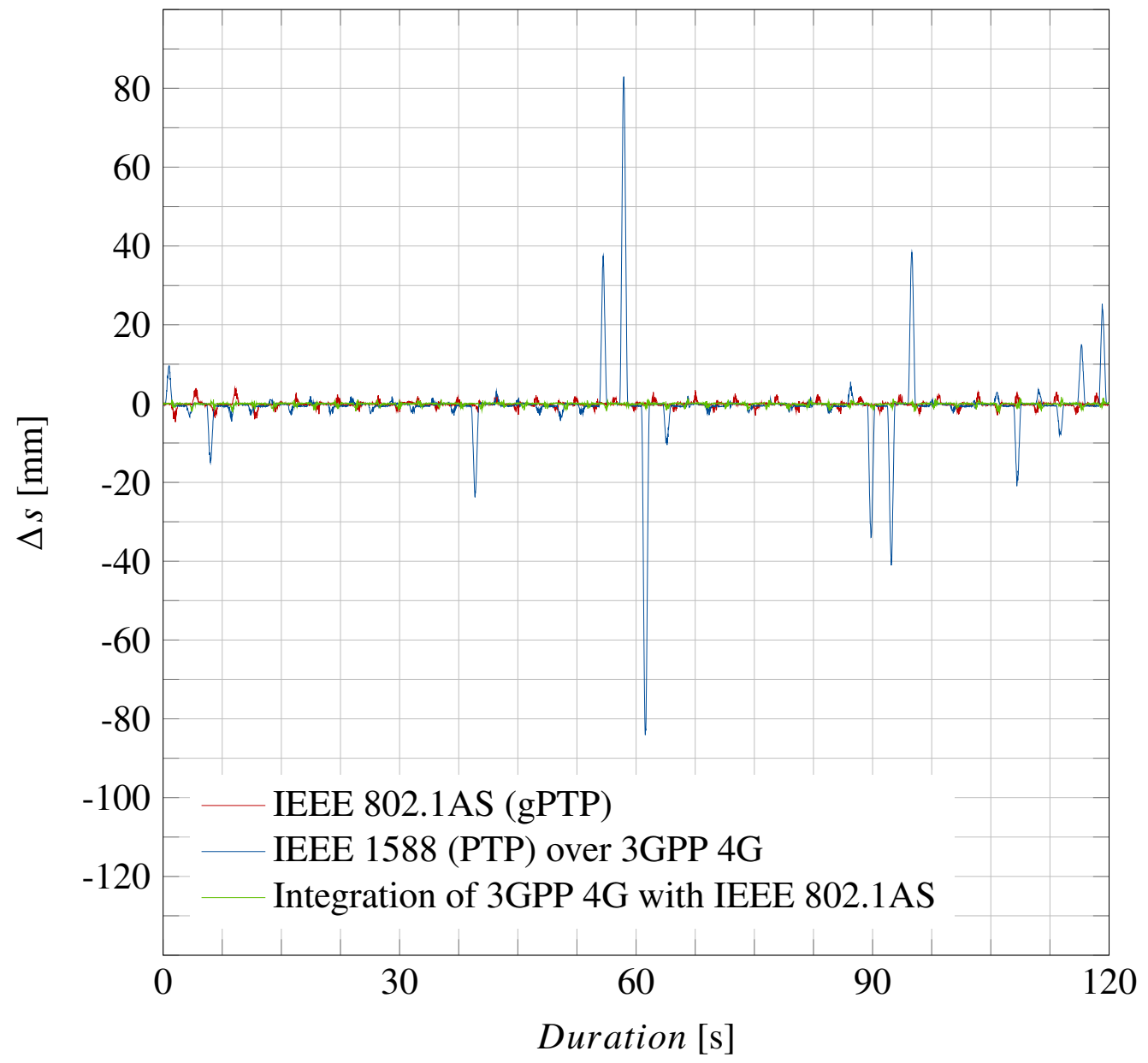

Figure 10. Visualization of the different measurements for the evaluation of our concept in a common plot.

## VI. Conclusion

In this paper we derived the synchronicity requirements from our target use case group, the cooperative work of mobile robots, identifying a synchronicity of $<$1ms as suitable.

In addition, we proposed a concept for the integration of IEEE 802.1AS time synchronization of TSN with 3GPP systems. Besides the application in 5G systems, this concept is also applicable for 4G.

Furthermore, we presented a discrete automation demonstrator and evaluated our approach based on three different measurements in combination with an open source 4G system. Using 4G as an example, the results show that our concept represents a significant improvement compared to the state of the art. Moreover, for this hardware configuration, which already has a high dynamic level, an accuracy for time synchronization of <1ms is sufficient. This means that for this setup the synchronization accuracy achieved with our concept by using 4G has the same performance as the more precise wireline synchronization using IEEE 802.1AS.

## References

[1] H. Kagermann, W.-D. Lukas, and W. Wahlster. (Apr. 2011). Industrie 4.0: Mit dem Internet der Dinge auf dem Weg zur 4. industriellen Revolution.

[2] M. Gundall, D. Reti, and H. D. Schotten, "Application of Virtualization Technologies in Novel Industrial Automation: Catalyst or Show-Stopper?", in *2020 IEEE 18th International Conference on Industrial Informatics (INDIN)*, IEEE, vol. 1, 2020, pp. 283–290.

[3] M. Gundall, C. Glas, and H. D. Schotten, "Introduction of an Architecture for Flexible Future Process Control Systems as Enabler for Industry 4.0", in *2020 25th IEEE International Conference on Emerging Technologies and Factory Automation (ETFA)*, IEEE, vol. 1, 2020, pp. 1047–1050.

[4] IEEE802 TSN Task Group, *TSN Standard*, 2020. [Online]. Available: https://1.ieee802.org/tsn/.

[5] J. L. Messenger, "Time-Sensitive Networking: An Introduction", *IEEE Communications Standards Magazine*, vol. 2, no. 2, pp. 29–33, Jun. 2018. DOI: 10.1109/MCOMSTD.2018.1700047.

[6] TACNET 4.0. [Online]. Available: http://www.tacnet40.com.

[7] M. Gundall, J. Schneider, H. D. Schotten, M. Aleksy, D. Schulz, N. Franchi, N. Schwarzenberg, C. Markwart, R. Halfmann, P. Rost, D. Wübben, A. Neumann, M. Düngen, T. Neugebauer, R. Blunk, M. Kus, and J. Grießbach, "5G as Enabler for Industrie 4.0 Use Cases: Challenges and Concepts", in *2018 IEEE 23rd International Conference on Emerging Technologies and Factory Automation (ETFA)*, vol. 1, Sep. 2018, pp. 1401–1408. DOI: 10.1109/ETFA.2018.8502649.

[8] C. Mannweiler, B. Gajic, P. Rost, R. S. Ganesan, C. Markwart, R. Halfmann, J. Gebert, and A. Wich, "Reliable and Deterministic Mobile Communications for Industry 4.0: Key Challenges and Solutions for the Integration of the 3GPP 5G System with IEEE", in *Mobile Communication - Technologies and Applications; 24. ITG-Symposium*, May 2019, pp. 1–6.

[9] M. Tao, "Collaborative robotics market to account for 30 per cent of total robot market by 2027, says report", *URL:https://roboticsandautomationnews.com/2019/12/05/collaborative-robotics-market-to-account-for-30-per-cent-of-total-robot-market-by-2027-says-report/27217/*, 2019.

[10] "IEC 62541-1, OPC Unified Architecture - Part 1: Overview and concepts", 2010.

[11] "IEC 62541-14, OPC Unified Architecture - Part 14: PubSub", 2019.

[12] J. Pfrommer, A. Ebner, S. Ravikumar, and B. Karunakaran, "Open Source OPC UA PubSub over TSN for Realtime Industrial Communication", Jul. 2018. DOI: 10.1109/ETFA.2018.8502479.

[13] E. Tiedemann, "5G: It's Here! More is Coming!", 2019. [Online]. Available: https://icc2019.ieee-icc.org/sites/icc2019.ieee-icc.org/files/ICC19KN_G-052319-Qualcomm.pdf.

[14] 5G Alliance for Connected Industries and Automation (5G ACIA), "5G Non-Public Networks for Industrial Scenarios", Jul. 2019.

[15] P. Popovski, K. F. Trillingsgaard, O. Simeone, and G. Durisi, "5G Wireless Network Slicing for eMBB, URLLC, and mMTC: A Communication-Theoretic View", *IEEE Access*, vol. 6, pp. 55 765–55 779, 2018.

[16] P. Rost, D. Chandramouli, and T. Kolding, "5G plug-and-produce", Apr. 2020. [Online]. Available: https://onestore.nokia.com/asset/207281.

[17] 3GPP, "TS 24.535 5G System (5GS); Device-Side Time Sensitive Networking (TSN) Translator (DS-TT) to Network-Side TSN Translator (NW-TT) protocol aspects; Stage 3", Jul. 2020. [Online]. Available: http://www.3gpp.org/DynaReport/24535.htm.

[18] ——, "TS 23.501 System architecture for the 5G System (5GS)", Mar. 2020. [Online]. Available: http://www.3gpp.org/DynaReport/23501.htm.

[19] A. Omri, M. Shaqfeh, A. Ali, and H. Alnuweiri, "Synchronization Procedure in 5G NR Systems", *IEEE Access*, vol. 7, pp. 41 286–41 295, 2019.